\documentclass[a4paper]{mem_sissa}
\usepackage{natbib}
\usepackage{graphicx}
\usepackage[a4paper]{hyperref}
\begin{document}
   \title{The VSNG Project: variable stars in nearby galaxies
\thanks{Based on data collected with the ESO and HST telescopes}
}

   \author{Gisella Clementini \inst{1} 
}

   \offprints{G. Clementini}
\mail{INAF - Osservatorio Astronomico,
via Ranzani 1, 40127 Bologna, Italy}

   \institute{INAF - Osservatorio Astronomico,
via Ranzani 1, 40127 Bologna, Italy
             }

   \abstract{We  describe the VSNG Project, a collaboration
   between researchers of Bologna, Padova, Napoli and Merate Observatories, to
   study variable stars in a number of galaxies in the Local Group. 
We briefly review 
   results obtained for the galaxies analyzed so far.
   }
   \authorrunning{G. Clementini}
   \titlerunning{The VSNG project}
   \maketitle
%

\section{Introduction}

Variable stars are fundamental tools for the definition of the astronomical distance
scale and for tracing stellar populations in  galaxies.
\par\noindent
The VSNG project: Variable Stars in Nearby Galaxies, is a collaboration 
between researchers of Bologna, Padova, Napoli and Merate Observatories, 
who have joined their efforts and their
skills to make a systematic study of the variable star populations  
in a number of Local Group (LG) galaxies. 
Started a few years ago, the project has already gathered data on 6 
galaxies of different morphological type, namely: the Large Magellanic Cloud 
(LMC), 
Fornax, Leo I, Phoenix, NGC 6822
and M31. Complete studies have been carried out or are in progress for  
the LMC (Bragaglia et al. 2001, Clementini et al.
2003a),
Leo I (Held et al. 2001), 
and NGC 6822 (Clementini
et al. 2003b). Analysis has just been started in Fornax.
Feasibility projects have been 
conducted in Phoenix, and M31 (Clementini et al. 2001). 
Many people are involved in the above studies. 
Participation and co-authorship is described for each 
specific subproject.
\par\noindent
The different parts of the VSNG project 
are handled by the collaborating teams according to 
their specific skills and  expertise.  Different approaches and most suited 
data reduction and analysis strategies are 
adopted, depending on the available data:
\begin{itemize}
\item Wide Field Imager (WFI) and Very Large Telescope (VLT)
 photometric time series data for 
Leo I, Fornax, Phoenix and NGC\,6822 are handled in Padova by 
E.V. Held and collaborators (L. Rizzi, Y. Momany, I. Saviane), 
who developed dedicated software (Wide Field Padova Reduction Package: 
WFPRED) and automatic reduction procedures Rizzi \& Held (2003, in 
preparation).
PSF photometry 
is performed with  
DAOPHOT and ALLFRAME (Stetson 1994) 
\item variable stars identification and study is done in Bologna by 
G. Clementini and her collaborators (L. Baldacci, M. Maio, F. Matonti, in 
particular) using
both private dedicated software (VARFIND, GRphycal Analyzer of 
TIme Series: GRATIS), and image subtraction
techniques (ISIS 2.1, Alard 2000). 
\item spectroscopic and photometric data of the LMC subproject have 
been handled jointly by people in Bologna and
Padova (A. Bragaglia, G. Clementini, M. Maio, E. Taribello, E. Carretta, 
R.G. Gratton) 
\item HST data for the M31 Globular Clusters are reduced with 
ROMAFOT (Buonanno \& Iannicola 1989). Identification of the candidate RR Lyrae stars
is done through comparison with template light curves.
\item theoretical pulsational models are provided by the Napoli team
(M. Marconi).
\end{itemize}

   \begin{figure*}
   \centering
   \resizebox{\hsize}{!}{\rotatebox[]{0}
{\includegraphics{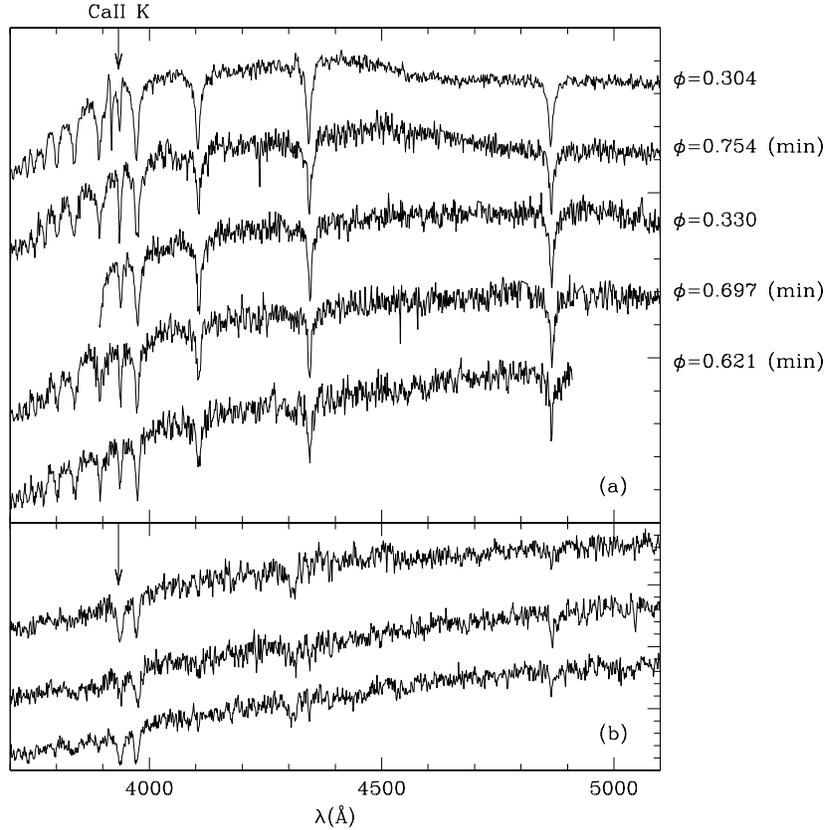}}}
   \caption{Spectra of RR Lyrae (panel {\it a}), and clump stars 
(panel {\it b})
in the bar of the LMC, obtained with FORS at UT1. 
For the RR Lyrae stars we label phases at which spectra were acquired.}
              \label{FigGam}%
    \end{figure*}

   \begin{figure*}
   \centering
   \resizebox{\hsize}{!}{\rotatebox[]{0}
{\includegraphics{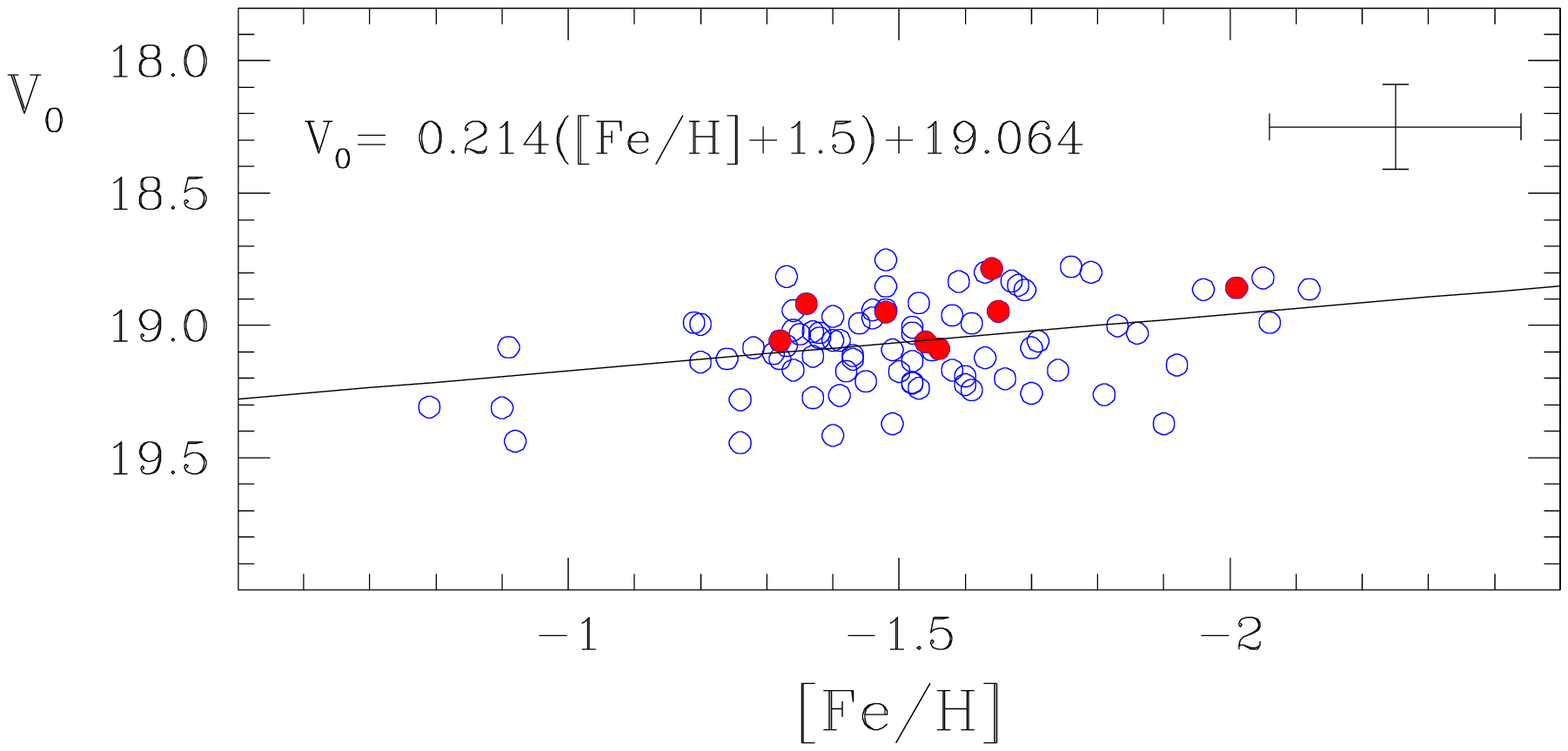}}}
   \caption{Luminosity-metallicity relation defined by 
RR Lyrae stars in the LMC. Filled (red) dots mark 
the double-mode pulsators in our sample (from Clementini et al. 2003a).}
              \label{FigGam}%
    \end{figure*}

\section {The Large Magellanic Cloud}
Involved people: G. Clementini, A. Bragaglia, E. Carretta, 
L. Di Fabrizio, R.G. Gratton, M. Marconi, M. Maio, E. Taribello
\medskip\noindent

The LMC is the first step of the astronomical distance scale. Many different
distance indicators coexist in the LMC, thus allowing a direct comparison of
the distance scale they provide. We observed 2 fields close to the bar of the 
LMC with the 1.5 m Danish telescope at ESO (La Silla) and obtained 
 $B$,$V$ light curves for 152 variables, among which 125 are RR Lyrae stars
(see contribution by Maio et al. in these Proceedings). 
Reddening in the two areas was estimated from the pulsational properties
of the RR Lyrae stars.
Low 
resolution spectroscopy (R$\sim$815) was obtained in 2001 with FORS at the
VLT for a hundred of the RR Lyrae stars, and for about 300 clump stars. 
Examples of our FORS1+UT1 spectra for RR Lyrae and 
clump stars in the LMC are shown in Figure~1. 
Metal abundances
for the RR Lyrae stars were derived using a revised version of 
the $\Delta S$
technique (Preston 1959), as fully described in Gratton et al. (2003, in 
preparation).
The average metal abundance of the sample is $<$[Fe/H]$>$=$-1.48$ ($\sigma$=0.29).
The new metal abundances and the apparent average luminosities of the
RR Lyrae stars were used to derive the following luminosity-metallicity 
relation:
$$<V_0(RR)> = 0.214 \times({\rm [Fe/H]} 
+ 1.5) + 19.064$$

This relation is shown in Figure~2. 
The dereddened average luminosity of the RR Lyrae stars is:
~~$<V(RR)_0>$=19.06$\pm$0.06 at [Fe/H]=$-1.5$. When combined with the
absolute magnitude of RR Lyrae stars provided by the Main Sequence Fitting 
of Galactic Globular clusters (Gratton et al. 2002) this value gives 
$\mu_{LMC}$=18.45$\pm 0.09$.
\par\noindent
A re-analysis of the distance moduli  derived for the LMC using various 
different 
distance indicators and techniques and the results of our LMC photometric
and spectroscopic study leads to $\mu_{LMC}$=18.515$\pm 0.085$, with no
longer dichotomy existing between short and long LMC distance scales 
Clementini et al. (2003a).
 
\section {Fornax and Leo I}
Involved people: G. Clementini, E.V. Held, E. Poretti, L. Rizzi,  
L. Baldacci, L. Di Fabrizio, M. Maio, F. Matonti, Y. Momany, L.
I. Saviane
\medskip\noindent  

Time series $B$, $V$ data for the dwarf spheroidal galaxies 
Leo I and Fornax were obtained with the Wide Field Imager of the 
 2.2 m MPG/ESO telescope.
\par\noindent
Data reduction of the Fornax data (18 $V$ and 62 $B$ frames covering about
1/4 of the galaxy) is in progress. Variable stars identification 
has been
successfully performed with ISIS2.1 on one of the 8 CCDs of the WFI mosaic.
The search will
be extended soon to the other CCDs.

For Leo I we have 40 $V$, 22 $B$ and 5 $I$ WFI frames 
covering the body of the galaxy and its surrounding fields. 
Data have been fully
reduced using the Padova pipeline for WFI frames. This consists
of the following steps:
\par\noindent
(a) bias subtraction and flat-fielding of the data with MSCRED (Valdes 1997), a public package 
 available into IRAF
\par\noindent
(b)reduction with WFPRED. This
software, developed within the IRAF environment by E.V. Held and L. Rizzi,  
allows
to make: 
\begin{itemize}
\item photometry of Landolt standard stars
\item astrometry of the science frames in automatic way to remove 
geometric distortions and generate the master catalogues
\item coaddition of dithered images
\end{itemize}
Due to technical problems, our time series
images of Leo I were not aligned, and astrometry was used to trim the whole 
set of images to a common intersection region, and to remove geometrical
distortions
\par\noindent
(c) photometry of the bias-subtracted, flat-fielded images with 
 ALLFRAME. The package 
 requires one PSF for each exposure for each CCD of the mosaic.
The 496 PSF's needed for our 62 WFI images of Leo I 
were generated through a fully automatic algorithm developed in 
Padova. ALLFRAME was then run on the pretreated images. The package produces a 
master object list by finding stars on the stack of all images, and 
performs PSF fitting photometry on the individual images using the 
master list. For each star the program returns both the average magnitude and a
record listing all measurements on the single frames.

Catalogues were then moved to Bologna to perform the identification 
of the variable stars and the period
definition.
Variables in Leo I were first identified in the three CCD's 
hosting the main body of the galaxy using the program VARFIND.
VARFIND is a private software developed at the Bologna Observatory 
by P. Montegriffo, which allows to pick up candidate variable 
stars from the scatter diagram of the 
time series measurements. 
The code is interactively linked to GRATIS, 
a private software which allows to perform period search and
study of single and multi-mode periodicities. GRATIS has been developed at the Bologna 
Observatory by P. Montegriffo,
G. Clementini and L. Di Fabrizio.
\par\noindent 
For the first time RR Lyrae stars were detected in Leo I
(Held et al. 2001), thus proving that the galaxy started forming stars at
an early epoch.
The distance modulus of Leo I derived from the luminosity level of the
Horizontal Branch traced by the RR Lyrae stars is:
$\mu_{Leo I}$=22.04$\pm 0.14$ (Held et al. 2001).
A deeper search of variable stars has been performed on the whole mosaic 
of the
8 CCDs using the ISIS2.1 package. Results are presented in
Clementini et al. (2003c, in preparation).

   \begin{figure*}
   \centering
   \resizebox{\hsize}{!}{\rotatebox[]{0}
{\includegraphics{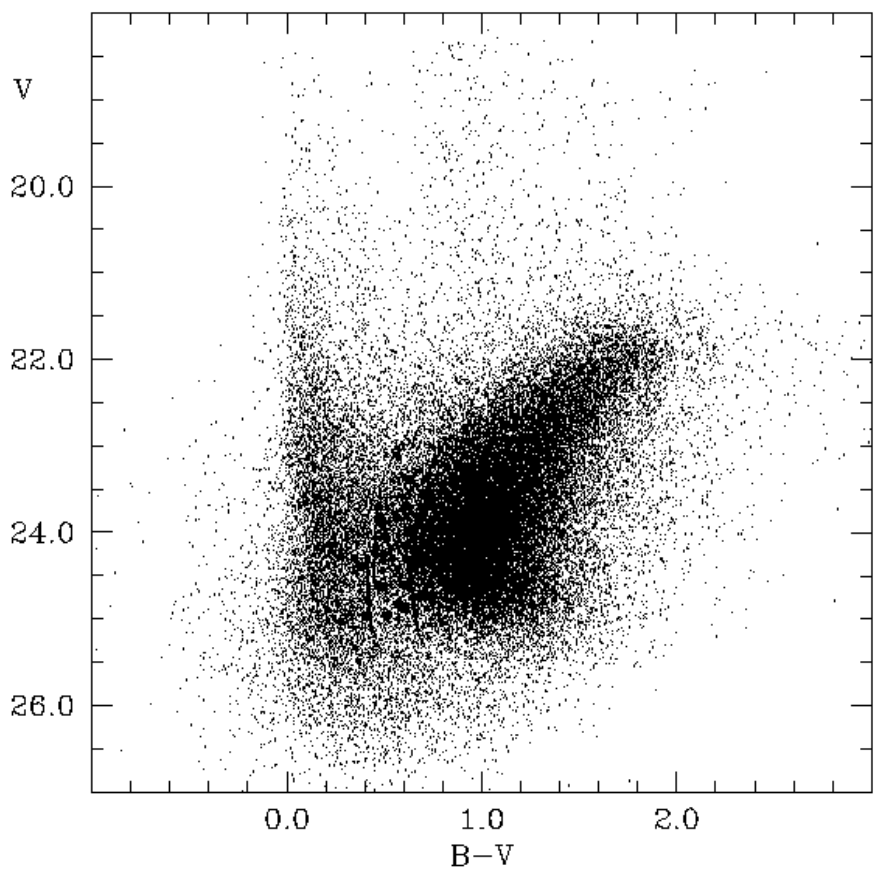}}}
   \caption{Color-magnitude diagram of NGC 6822 
(from Clementini et al. 2003b)
showing the location 
of the newly discovered RR Lyrae (filled circles) and 
Anomalous Cepheids (crosses) variable stars.
Solid lines show the edges of the RR Lyrae instability strip of M 3
(Corwin \& Carney 2001), dashed lines are the the instability strip
boundaries for 
1.5 M$_{\odot}$ models (Bono et al. 1997).}
              \label{FigGam}%
    \end{figure*}

\section {Phoenix and NGC 6822}
Involved people: G. Clementini, E.V. Held, L. Baldacci, L. Di Fabrizio, 
Y. Momany, L. Rizzi, I. Saviane
\medskip\noindent  

Both Phoenix and NGC6822 were observed with the VLT.
\par\noindent
Phoenix time series arise from archival data and consist of: 10 $V$, 7 $B$ 
and 4 $I$ frames. 
The color magnitude diagram of Phoenix shows a very well defined HB 
(Held et al. 2003, in preparation) 
at raw magnitude about $v \sim 15.5$, and 
a well defined RR Lyrae gap around $b-v \sim 0.8$. Indeed, the scatter 
diagram obtained with VARFIND clearly shows large  $\sigma$'s corresponding
to the RR Lyrae {\it finger} at
$v \sim 15.5$. 
Although we only have a few frames for Phoenix, using the scatter diagram and 
playing with data in the three different bands we were able (i) to identify 
candidate RR Lyrae stars,  and (ii) to build up multicolor light curves
that can then be compared with RR Lyrae template light curves.

NGC6822 time series consist of 36 $V$, 11 $B$, and 4 $I$ VLT frames.
Variable star identification performed with ISIS2.1
led to the discovery of a conspicous number of RR Lyrae stars and 
Anomalous Cepheids in the galaxy.
A  more thorough description of the variable star identification technique 
and results is
provided in  Baldacci et al. (2003, these Proceedings). 
Figure~3 shows the location of 
 the newly discovered RR Lyrae (filled circles) and 
Anomalous Cepheid (crosses) stars in the 
calibrated color-magnitude diagram of NGC 6822
obtained with ALLFRAME.
Solid lines show the edges of the RR Lyrae instability strip of M 3
(Corwin \& Carney 2001), dashed lines are the instability strip
boundaries for 
1.5 M$_{\odot}$ models (Bono et al. 1997).
A preliminary estimate of the 
average luminosity of the RR Lyrae stars identified in 
NGC 6822 is $<V(RR)>$=24.61$\pm 0.14$, leading to 
a distance modulus of: 
$\mu_{NGC 6822}$=23.35$\pm 0.15$, for E(B-V)=0.24, [Fe/H]=$-1.8$ and
M$_{V}(RR)$= 0.57 mag (Clementini et al. 2003b).

\section {M31}
Involved people: G. Clementini, L. Federici, C. Corsi, C. Cacciari, 
M. Bellazzini, H.A. Smith.
\medskip\noindent  

A feasibility study was performed 
on archival HST data of 4 Globular
clusters (GCs) in the Andromeda galaxy.
We used 4 F555W and 4 F814W observations of four M31 
GCs to identify RR Lyrae stars.
Candidate
variables were selected from 
the VARFIND scatter diagrams for the 4 clusters.
Time series data for each candidate variable
were then compared with template light curves 
of RR Lyrae stars in the Galactic globular cluster M3 (Carretta et al. 1998).
We were able to identify 2, 4, 11 and 8 such candidate RR Lyrae stars in 
G11, G33, G64 and G322, respectively, as fully described in Clementini et al.
(2001).


\begin{acknowledgements}
      Part of this work was supported by the MURST Cofin98 under the projects 
      "Stellar Evolution" and "Processing of large astronomical 
images",  and by MURST - Cofin00 under the project 
      "Stellar observables 
of cosmological relevance"     
\end{acknowledgements}

\bibliographystyle{aa}

\end{document}